\documentclass[aps,prl,twocolumn,showpacs]{revtex4}
\usepackage{graphicx}

\begin{document}

\title{Appearance of an inhomogeneous superconducting state in
La$_{0.67}$Sr$_{0.33}$MnO$_{3}$-YBa$_{2}$Cu$_{3}$O$_{7}$-La$_{0.67}$Sr$_{0.33}$MnO$_{3}$
trilayers}

\author{K. Senapati}
\author{R. C. Budhani}
\email{rcb@iitk.ac.in}
%\homepage[]{Your web page}
%\thanks{}
%\altaffiliation{}
\affiliation{Department of Physics, Indian Institute of Technology
Kanpur, Kanpur - 208016, India}

\date{\today}

\begin{abstract}
An experimental study of proximity effect in
La$_{0.67}$Sr$_{0.33}$MnO$_3$ - YBa$_2$Cu$_3$O$_7$ -
La$_{0.67}$Sr$_{0.33}$MnO$_3$ trilayers is reported. Transport
measurements on these samples show clear oscillations in critical
current (I$_c$) as the thickness of La$_{0.67}$Sr$_{0.33}$MnO$_3$
layers (d$_F$) is scanned from $\sim$ 50 {\AA} to $\sim$ 1100
{\AA}. In the light of existing theories of
ferromagnet-superconductor (FM-SC) heterostructures, this
observation suggests a long range proximity effect in the
manganite, modulated by it's weak exchange energy ($\sim$ 2 meV).
The observed modulation of the magnetic coupling between the
ferromagnetic LSMO layers as a function of d$_F$, also suggests an
oscillatory behavior of the SC order parameter near the FM-SC
interface.

\end{abstract}

% insert suggested PACS numbers in braces on next line
% insert suggested keywords - APS authors don't need to do this
%\keywords{}

\maketitle

The interplay between superconductivity (SC) and ferromgnetism
(FM) in FM-SC heterostructures leads to some interesting physical
phenomena, one of which is the observed non-monotonic dependence
of transition temperature T$_c$ \cite{Wong, Strunk, Jiang, Lazar,
Obi, Zhao} and critical current I$_c$ \cite{Ryazanov, Blum} on the
thickness of ferromagnetic layer (d$_F$). Theoretically, such
systems are treated as a boundary value problem, solving the
Usadel equations \cite{Usadel} for anomalous Green's functions on
both sides of the FM-SC interface. The pioneering work of Radovic
et al. \cite{Radovic} using this formalism, has successfully
reproduced the non-monotonic behavior of T$_c$ in FM-SC
multilayers \cite{Wong, Strunk, Jiang} and SC-FM-SC trilayers
\cite{Strunk}. Similar calculations by Buzdin et al. \cite{Buzdin}
have revealed an oscillatory nature of Josephson current across a
ferromagnetic spacer. However, the Radovic-Buzdin (RB) theory,
which relies on competing ``0'' and ``$\pi$'' phase coupling
between adjacent superconducting layers to explain the
non-monotonic nature of T$_c$ and I$_c$, cannot be applied to
systems where there is only one SC layer in contact with a
ferromagnetic film such as the FM-SC-FM trilayer and FM-SC bilayer
structures. Another restricted point of the RB theory is the
assumption of perfect transparency of the FM-SC interface.
Recently, these issues have been addressed using more realistic
boundary conditions \cite{Khusainov, Tagirov, Fominov}. In
general, the microscopic basis of these theories is the formation
of a Larkin-Ovchinnikov-Fulde-Ferrel (LOFF)  \cite{LO, FF} like
inhomogeneous SC-state at the FM-SC boundary \cite{Demler}.

On the experimental scenario, there are increasing number of
reports \cite{Wong, Strunk, Jiang, Lazar, Obi, Zhao} on the
oscillating nature of T$_c$(d$_{_F}$) although negative results
\cite{Koorevaar} and different interpretations \cite{Muhge} have
also been reported in some cases. A more sensitive way of
addressing this issue is the measurement of critical current
I$_c$(d$_{_F}$) through FM-SC interfaces. Such studies
\cite{Ryazanov, Blum} have unambiguously established the existence
of an oscillating order parameter. However, these results are
explained on the basis of $\pi$-phase coupling between two
superconducting layers. In this paper, we report the observation
of oscillating critical current in FM-SC-FM trilayer structures
where the concept of $\pi$-coupling does not apply altogether.
Unlike the itinerant ferromagnet-weak coupling BCS superconductor
based structures, the constituents in the present case are exotic,
showing localized spin ferromagnetism and a highly anisotropic
superconducting order parameter. This first-time observation of an
oscillating I$_c$ in such a system is remarkable.

\begin{figure}[b]
\vskip -1cm \abovecaptionskip 0cm
\includegraphics[width=7cm]{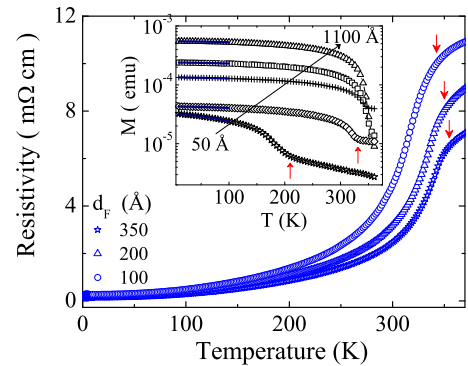}%
 \caption{(a) Resistivity ($\rho \left( {T} \right))$ of LSMO films deposited on
STO in the temperature range of 2 K - 370 K. Thickness of the
films varies from 100 {\AA} to 350 {\AA}. Inset: The 1000 Oe
field-cooled magnetization of single layer LSMO films of thickness
ranging from 50 {\AA} to 1100 {\AA}. In all cases the magnetic
field was applied in the plane of the film. The solid lines are
fits to the Bloch relation (see text for details). Curie
temperatures have been marked by the arrows.}
 \end{figure}
We have studied a series of high quality trilayer structures in
FM-SC-FM geometry with a $\sim$100 {\AA} superconducting
YBa$_2$Cu$_3$O$_7$ (YBCO) layer sandwiched between ferromagnetic
La$_{0.67}$Sr$_{0.33}$MnO$_3$ (LSMO) layers, prepared by pulsed
laser ablation on single crystal SrTiO$_3$ substrates. Thickness
of the LSMO layers (d$_F$) were varied from $\sim$50 {\AA} to
$\sim$1100 {\AA}. Details of film growth are described elsewhere
\cite{Senapati}. The suitability of the CMR manganite-high T$_c$
superconductor combination for epitaxial growth is also
well-established in the literature \cite{Jacob, Goldman,
Habermeir, Pena}.

The magnetic nature of the LSMO layers as a function of thickness
was established from transport and magnetization measurements.
Fig. 1 shows the resistivity ($\rho(T)$) of few representative
LSMO thin films in the temperature range of 2 K and 370 K.
Resistivity of these films at room temperature is low ($\sim $ 2
m$\Omega $cm), and remains metallic down to 2 K. The paramagnetic
metallic phase above the Curie temperature (T$_{Curie}$)
\cite{Urushibara} which transits to a ferromagnetic metallic phase
at T $<$ T$_{Curie}$, is clearly identifiable in all films. The
ordering temperature acquires the near bulk value ($\sim $ 350 K)
in films thicker than 200 {\AA}, while thinner films show a slight
drop in T$_{Curie}$, consistent with earlier measurements on
ultrathin LSMO films \cite{Sun}. We have estimated the
ferromagnetic exchange energy by fitting the 1000 Oe field-cooled
M$_s$(T) measurements (shown in Fig. 1) to the Bloch relation
$M_{s} (T) / M_{s} (0) = 1 - AT^{3 / 2}$. Here $A = (C / S)(k_{B}
/ 2E_{ex}S)^{3 / 2}$, where `S' is the total spin per Mn ion in
LSMO and $C$ is the Bloch constant with a value 0.059 for a cubic
lattice \cite{Kittel}. All trilayer samples were rigorously
checked for simultaneous occurrence of superconductivity and
magnetism, using magnetization and transport measurements
\cite{Senapati}.
\begin{figure}[b]
\vskip -.5cm \abovecaptionskip -.4cm
 \includegraphics[width=8cm]{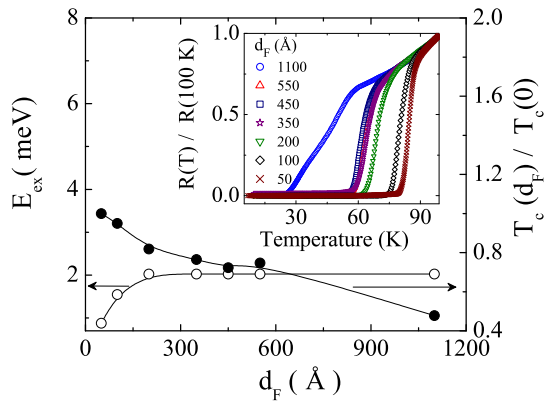}%
 \caption{ Superconducting transition temperature (T$_c$)
 of trilayers, normalized with respect to the T$_c$ of a 100 {\AA} YBCO in LSMO-YBCO-LSMO trilayer,
 is plotted (on right y-axis) with the thickness (d$_F$) of LSMO boundaries.
 The exchange energies of corresponding single layer LSMO films (calculated from the
 fittings in the inset of Fig. 1) is plotted on the left y-axis.
 The solid lines are only guides to the eyes.
 Inset shows the resistive transitions of the trilayers into the
 superconducting state as a function of temperature.}
 \end{figure}

The superconducting transitions as seen in $\rho(T)$ measurements
on various trilayers are presented in the inset of Fig. 2. The
one-step transitions seen in this inset exclude the possibility of
any metallurgical activities between LSMO and YBCO, which would
otherwise lead to the formation of a degraded phase of YBCO at the
interface, with lower T$_c$. The transition temperature
T$_c$(d$_F$) of the trilayers normalized with respect to the T$_c$
of a trilayer with only 50 {\AA} LSMO on both sides of YBCO is
plotted in Fig. 2 as a function of d$_F$. The T$_c$(d$_F$) has
been defined as the temperature at which the sample resistance
reaches half the extrapolated normal state resistance. Fig. 2 also
shows the variation of exchange energy (E$_{ex}$) extracted from
the M(T) data of Fig. 1 with d$_F$. The calculated value of
E$_{ex}$ in the thick limit ($\sim$ 2 meV) is in good agreement
with the results obtained directly from ferromagnetic resonance
measurements on similar films \cite{Golosovsky}. The decay of
T$_c$ with d$_F$ in Fig. 2 is primarily monotonic except for the
appearance of a plateau in the neighborhood of d$_F \sim$ 450
{\AA}. The absence of oscillations in the T$_c$(d$_F$) curve
suggests a limited transparency of the FM-SC interface. In spite
of the near perfect lattice matching between LSMO and YBCO some
uncontrollable factors, like the Fermi-velocity mismatch between
the two materials in contact may lead to a smearing of the
T$_c$(d$_F$) oscillations\cite{Tagirov}.

\begin{figure}[t]
\vskip -.5cm \abovecaptionskip -.4cm
 \includegraphics[width=8.5cm]{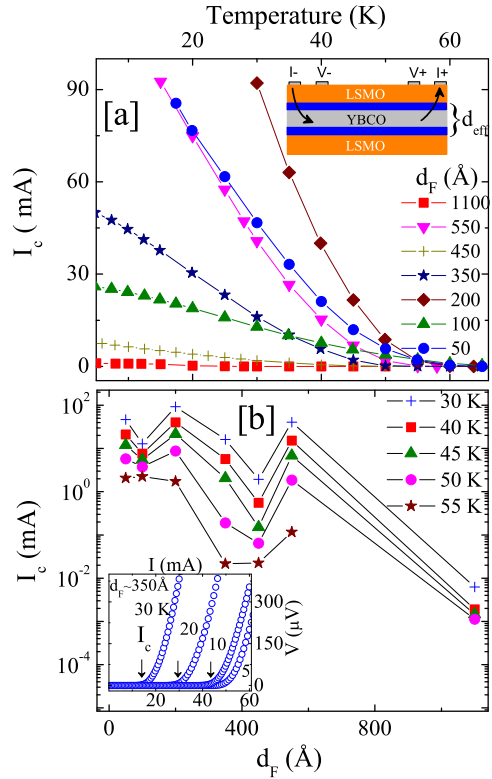}%
 \caption{(a) In-plane critical current (I$_c$) of LSMO-YBCO-LSMO trilayers
plotted as a function of temperature. Inset is a sketch of the
measurement geometry. The shaded portions at the YBCO-LSMO
interfaces are the inhomogeneous superconducting regions which
contribute to the overall critical current of the system. (b) The
data of panel `a' have been re-plotted as I$_c$(d$_F$) isotherms
at several temperatures. Inset shows the IV curves of a trilayer
with d$_F \sim$350 {\AA} at temperatures 5, 10, 20, and 30 K.
Arrows indicate the critical current I$_c$.}
 \end{figure}

The critical current I$_c$(d$_F$) was measured in a standard
four-probe geometry, as shown in a sketch in Fig. 3(a). Although
in the normal state both LSMO and YBCO layers act as parallel
conducting channels for the current, in the superconducting state
current is preferentially directed into the YBCO. However, owing
to the small thickness of the superconducting channel in our
trilayers and the induced superconducting order at the boundary,
the proximally important interface region of the FM layers (shaded
portion at the LSMO-YBCO interface, shown in the sketch of Fig.
3(a)) also contributes to the flow of supercurrent. Clearly, as
the YBCO thickness is fixed in all cases, magnitude of I$_c$ is
expected to reflect the relative amplitude of the pair-wave
function in different samples. The critical current I$_c$ has been
extracted from the measurements of current-voltage
characteristics, as shown in the inset of Fig. 3 (b) for a
trilayer with d$_F \sim$350 {\AA}. In Fig. 3(a) we show the I$_c$
of all trilayers as a function of temperature. The behavior of
I$_c$ is clearly non-trivial as the thickness of LSMO boundaries
in these heterostructures is varied. The same data have been
plotted as isothermal curves at several temperatures as a function
of d$_F$ in Fig. 3(b). The behavior of I$_c$ is most certainly
oscillatory with an average period of $\sim $250 {\AA} and more
than an order of magnitude change in current between the maxima
and minima. Theoretically, this period corresponds to the distance
over which the induced pair wave-function changes it's phase by
$\pi$ according to the relation ($\pi \hbar v_F /E_{ex}$)
\cite{Demler}. Assuming a LOFF-like picture for the current
situation and using the measured exchange energy (2 meV), we
obtain a Fermi velocity v$_F \sim$ 2.4$\times$10$^6$ cm/sec, which
is somewhat different from the value ($\sim$ 7.4$\times$10$^7$
cm/sec) derived from band structure calculations \cite{Pickett}.
This discrepancy might be a reflection of the large uncertainty
involved in determining the value of v$_F$ for CMR manganites from
band structure calculations, due to strong hybridization effects
of Mn-$d$ and O-$p$ bands.

\begin{figure}
\vskip -.5cm \abovecaptionskip -.4cm
 \includegraphics[width=7cm]{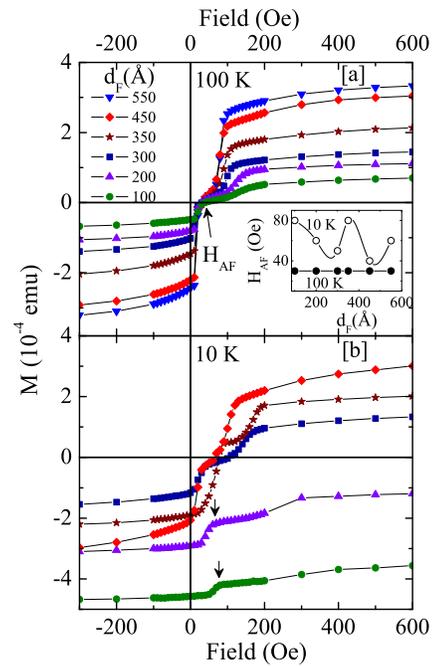}%
 \caption{Panels `a' and `b' show the last two quadrants of
 zero-field-cooled hysteresis curves at 100 K and 10 K respectively.
 All measurements were carried out with a field-in-plane geometry.
 The inset in panel `a' compares the antiferromagnetic coupling field
 H$_{AF}$ (indicated by arrows in some cases) at temperatures 10 K and 100 K, as a function of
 the thickness of LSMO boundary (d$_F$).
 The solid lines in the inset are only guides to the eyes.}
 \end{figure}

To further verify the oscillating nature of I$_c$(d$_F$), we
conducted dc-magnetization measurements where diamagnetic
supercurrents are intrinsically generated inside the YBCO layer in
response to the applied magnetic field. These measurements were
performed with an in-plane field geometry which produces the
screening currents along the cross section of the trilayers. The
diamagnetic moment of this induced current acts as an opposing
field which suppresses the effective magnetic field felt by the
LSMO layers. Therefore, a change in the induced current
(equivalently the diamagnetic moment) should be detectible from
the magnetic coupling behavior of the LSMO boundaries.
Zero-field-cooled magnetization measurements on our trilayer
samples revealed a clear region of antiferromagnetic coupling
between the moments of the top and the bottom LSMO layers at low
fields ($<$200 Oe), as manifested by a plateau in the
magnetization curve. Fig. 4(a) shows the last two quadrants of the
hysteresis loops measured at 100 K, where the YBCO is still in the
normal state. The antiferromagnetic coupling field (H$_{AF}$)
extracted from the M-H loops at 100 K is found to be the same
(30$\pm$5 Oe) for all samples. Panel (b) of Fig. 4 shows the
magnetization measured at 10 K. Here the ferromagnetic
contribution of the LSMO layers is superimposed on the strong
diamagnetic moment of YBCO. However, the plateau arising from
antiferromagnetic coupling between the LSMO layers is still
observable. Furthermore, in clear contrast to the data at 100 K,
the coupling field H$_{AF}$ in this case is oscillatory with
d$_F$, as shown in the inset of Fig. 4(a). The oscillatory
behavior appears to be a signature of the modulation of screening
critical currents. Most interestingly, the period of oscillation
in this case is found to be $\sim$200 {\AA}, which is close to the
period ($\sim$ 250 {\AA}) obtained earlier from transport
I$_c$(d$_F$). The large range of proximity effect seen here is
consistent with the results of Kasai et al. \cite{Kasai}, who have
reported a measurable supercurrent across YBCO-LSMO-YBCO trilayer
junctions with LSMO spacers of the order of 1000 {\AA}.

As already mentioned, the current observations can not be
explained on the basis of $\pi$-phase coupling, since here we have
only one superconducting layer. This difficulty has been addressed
by more recent theories \cite{Khusainov, Tagirov, Fominov},
predicting similar oscillations in heterostructures consisting of
a single superconducting layer. We, however, realize the
difficulty in mapping the current situation onto these theories
which have been developed assuming the s-wave symmetry of the
superconductors order parameter. On the other hand, there is
overwhelming experimental evidence for a d-wave pairing symmetry
in YBCO, with pair transport along the c-axis occurring only via
Josephson tunneling.  However, a few points independent of the
symmetry of the order parameter can be picked up for a qualitative
analysis. The non-monotonic changes in the superconducting
properties with d$_{_F}$ can be understood from the predicted
\cite{Izyumov_review} non-monotonic drop in the pair-amplitude at
the FM-SC interface, constrained by a maximum at the outer
boundary of the ferromagnet. When a node (minimum) of the pair
wave-function appears at the FM-SC interface, the Cooper pairs
entering the ferromagnet die quickly. On the other hand, an
antinode at the interface provides better chances of survival for
the Cooper pairs. Thus, the appearance of nodes and antinodes at
the interface should manifest as a minimum and maximum in
T$_c$(d$_{_F}$) and I$_c$(d$_{_F}$) curves.

The exact mechanism by which the supercurrent is continued as a
quasiparticle current in an adjacent ferromagnetic layer is not
known yet. However, the zero energy Andreev bound states, believed
to be the origin of zero bias conductance peaks (ZBCP) observed in
HTSC, might play a role here. Kasiwaya et al. \cite{Kasiwaya1,
Kasiwaya2} have shown that such bound states may lead to a
spontaneous quasiparticle current across a
ferromagnet-d$_{x^2-y^2}$-wave superconductor junction depending
on the phase of order parameter at the interface, when the
interface is perpendicular to the ab-plane. Interestingly, the
ZBCP is also seen in LSMO-YBCO junctions where the granularity of
the c-axis oriented YBCO leads to sampling of ab-plane Andreev
bound states \cite{Chen}.

In conclusion, we have observed clear oscillations in critical
current of LSMO-YBCO-LSMO trilayers as a function of the LSMO
thickness. The period of oscillation was found to be large
($\sim$200 {\AA}). This non-monotonic behavior appears to be a
manifestation of the LOFF-like oscillatory superconducting order
parameter near the FM-SC interface in the limit of weak exchange
energy (E$_{ex} << $ k$_B$T$_c$). The magnetic coupling behavior
of the LSMO boundaries also points towards similar results. To our
knowledge, this is the first observation of oscillatory critical
current as a function of d$_F$ in a manganite-cuprate
heterostructure.

\begin{acknowledgements}
This research has been supported by a grant from the Defense
Research {\&} Development Organization, Govt. of India. We thank
Prof. T. V. Ramakrishnan and Prof. Deepak Kumar for their critical
comments on this work.

\end{acknowledgements}

\end{document}